\documentclass{llncs}
\usepackage[T1]{fontenc}
\usepackage[latin9]{inputenc}
\usepackage[english]{babel}
\usepackage{epsfig}
\usepackage{nicefrac}
\usepackage{graphicx}
\usepackage{setspace}
\usepackage{multirow}
\usepackage{fancyvrb}
\usepackage{pst-node}
\usepackage{fullpage}


\begin{document}
\date{}

\title{Simulation of the Lattice QCD and Technological Trends in Computation}

\author{K. Ibrahim\inst{1} \and J. Jaeger\inst{2} \and Z. Liu\inst{6,7} \and L.N. Pouchet\inst{3,4} \and P. Lesnicki\inst{3,4} \and L. Djoudi\inst{2} \and D. Barthou\inst{2,3} \and F. Bodin\inst{1} \and
 C. Eisenbeis\inst{3} \and G. Grosdidier\inst{5} \and O. P\`ene\inst{7} \and P. Roudeau\inst{5}}

\institute{
Irisa/Inria, Rennes, \{kibrahim, bodin\}@irisa.fr 
\and
Universit\'e de Versailles St Quentin
\and
Alchemy team, Inria Saclay 
\and 
Laboratoire de Recherche en Informatique, Universit\'e de Paris XI
\and
Laboratoire de l'Acc\'el\'erateur Lin\'eaire, Universit\'e de
  Paris XI and IN2P3/CNRS
\and
LPSC, Universit\'e Joseph Fourier Grenoble 1, 
IN2P3/CNRS, INPG, 53 Av. des Martyrs, 38026 Grenoble
\and
Laboratoire de Physique Th\'eorique\footnote{Unit\'e Mixte de Recherche 8627 du Centre National de 
la Recherche Scientifique}
{Universit\'e de Paris XI, MPPI/CNRS}
}

\maketitle

\begin{abstract}
Simulation of Lattice QCD is a challenging computational problem. Currently,
technological trends in computation show multiple divergent models of
computation. We are witnessing homogeneous multi-core architectures, the use of
accelerator on-chip or off-chip, in addition to the traditional architectural
models. \\
 On the verge of this technological abundance, assessing the performance
trade-offs of computing nodes based on these technologies is of crucial
importance to many scientific computing applications. \\
 In this study, we focus on assessing the efficiency and the performance expected
for the Lattice QCD problem on representative architectures and we project the
expected improvement on these architectures and their impact on performance for
Lattice QCD. We additionally try to pinpoint the limiting factors for
performance on these architectures.

\end{abstract}

\section{\label{sec:Intro}Introduction}

Quantum chromodynamic (QCD) is the theory of strong interaction in the
domain of subnuclear physics. Lattice QCD (LQCD) is a numerical method
based on QCD first principles, the only one able to compute reliably
many quantities of high scientific relevance. It is based on a
discretization of space time and a Monte-Carlo method. The system
being an extremely complex one and the number of degrees of freedom
being of the order of a billion today, a number promised to increase
in the future, LQCD needs heavy, efficient and cheap enough computing
tools (hardware and software).

The goal of the calculation is to produce, according to a given
probability law resulting from the theory, a wide statistical sample of
``gauge configurations,'' each of which being a large file of complex
numbers. Although using efficient algorithms which will be described
in the next section, it requires a very large amount of computing power.

Simulation of this theory is one of the grand challenge problems in
part because of the small percentage of the computation load usually
being observed on most computing architectures.  Assessing the
performance and efficiency on new architectures, and based on
different algorithmic representation of this problem, is important to
get closer to the computational power needed for this problem. This
computation tends to have low utilization and efficiency on most
general-purpose computing facilities, leading to inefficient power
consumption and unrealistic demands on the number of required
computational nodes. So far building a special machine for
simulating the Lattice QCD problem has been a widely used
approach~\cite{Boyle01,Boyle05,Belletti06}. The motivation to build
specialized computing facilities, despite all the associated
overheads, is the enormous computational power needed in addition to
the special characteristics of the computation of the Lattice QCD.

For this problem, an optimal computing node should offer support for
complex arithmetic instructions, large register file, SIMD
instructions, software controlled cache management and balanced memory
bandwidth and communication bandwidth to the computational
power. Multiple successful balanced designs were historically build
for Lattice QCD~\cite{Boyle01,Boyle05,Belletti06}, but the overhead of
design and maintenance is usually high. Building out of commodity
components that best fit the problem characteristics is a very
attractive alternative, but it requires a careful analysis of the
problem, together with the analysis of the large spectrum of
architectural alternatives.

Currently, the driving forces for computer architecture push multiple
technologies with no clear convergence to performance/power overall
winner. We intend to explore these technologies, guided by the
computational requirements of the Lattice QCD. As it is extremely
difficult to exhaustively experiment all the emerging technologies, we
choose to focus on two groups of technologies:

\begin{itemize}
\item \textit{General-purpose homogeneous nodes:} we describe our
  efforts to assess the trade-offs of implementing the Lattice QCD
  computation on different architectures. We mainly focused on Itanium
  and Pentium architectures. These two architectures represent two
  major design alternatives in general purpose computing. The EPIC
  architecture for Itanium processor relies on the compiler in
  managing instruction parallelism, while the superscalar architecture
  for the Pentium processor relies on hardware management of
  instruction parallelism. Both architectures are deployed
  successfully to build highly scalable machines.
\item \textit{Heterogeneous computing nodes:} we investigate the use of
  specialized accelerators to improve the performance of a computing
  node. The first alternative we explored is the use of Intel Xeon
  processor assisted by a G80 NVIDIA graphic card as an accelerator, a
  heterogeneous multi-chip alternative. The second alternative we
  explored is based on integrating accelerators with the main
  processor on chip, providing a heterogeneous system-on-chip kind of
  architectures. A good representative of this architecture is the IBM
  Cell broadband engine.
\end{itemize}

The objective of this study is to compare future technologies prospect
for the simulation of Lattice QCD. We do not seek a generalized
comparative study of all future architectural trends; we target the
comparison based on the requirements for the simulation of the Lattice
QCD computations.

Our study reveals that the performance of the Lattice QCD computation
can be greatly improved using specialized accelerators. More
importantly, we predict that the imbalance of the computational power
to communication bandwidth for the Lattice QCD will remain an obstacle
for all the studied architecture. Efficient usage of the computational
power will rely heavily on the level of explicit resource management
that a particular hardware will offer.

The rest of this paper is organized as follows:
Section~\ref{sec:problem-LQCD} introduces the Lattice QCD problem and
its physical formulation. Section~\ref{sec:Singlenode} introduces
analysis of the performance of single node based on various
architectural alternatives. Section~\ref{sec:comparison-and-evolution}
discusses the needed improvements in the performance of the discussed
architectures and contrast it with their expected or planned
evolution. Section~\ref{sec:multinode} discusses the performance
impact of the communication architecture for a large scale system.
Section~\ref{sec:conclusions} concludes this work.

\section{\label{sec:problem-LQCD}The Problem of LQCD}
\def\bea{\begin{eqnarray}} \def\eea{\end{eqnarray}}
In Lattice QCD, the four-dimensional space-time continuum is simulated
by a four-dimensional lattice, of length respectively $X,Y,Z,T$ in the
four directions, with quark quantum fields on each lattice site and
gluon quantum fields represented by SU(3) matrices on each link
between these sites. SU(3) refers to $3\times 3$ unitary matrices of 
complex numbers of unit determinant. The 3-dimensional space in which these
matrices act is referred to as the space of the three quark ``colours". 
 The spinors are represented by four SU(3) vectors, each
composed of three complex variables. The calculation aims at computing
the average values of physical quantities, which are functions of
these fields, according to a probability distribution also depending
on the fields, and derived by a discretization procedure from the
basic QCD Lagrangian. This average is taken over the full space of all
the possible values of the fields. This space is known as the field
configuration space. The integration of quark fields is
done formally, leading to a complicated non-local probability
distribution: the determinant of the very large ``Dirac operator''
matrix, which depends only on the gluon fields. The probability law is
described by a complicated expression depending only on the gluon
fields i.e. the SU(3) matrices. We call gauge configuration a set of
SU(3) matrice matrices defined on all links. The probability law is thus
defined in the space of gauge configurations.

For large lattices the space of gauge configurations is a variety with
dimensionality of the order of billions. Only a Monte-Carlo method
allows such a huge calculation.  To estimate the average values of the
physical quantities we need representative samples of gauge
configurations (say about 5000) for every set of parameters, generated
according to the above-mentioned probability law. The Hybrid
Monte-Carlo (HMC) algorithm~\cite{Duane87}, or variants of it such as
the Polynomial HMC (PHMC), the Rational HMC (RHMC), is used to
generate these samples. This is a very heavy calculation. In the
following discussion, we will consider an HMC implementation achieved
by the ETMC collaboration~\cite{Urbach06,Urbach07}.

The run is decomposed into ``trajectories,'' which are indeed trajectories of  a
complex dynamical system depending on the SU(3) matrices. Each trajectory
leads from one gauge configuration to the next one of our sample. The
Hamiltonian of this system is devised in such a way as to generate gauge
configurations with a large enough probability. At the end of the trajectory  a
Metropolis test ensures the correct probability law. The trajectory is divided
into steps. After every step the gauge configuration is updated. During the
step, the gauge configuration stays unchanged. The algorithm manipulates
objects named ``Wilson spinors.'' One Wilson spinor is composed by a spinor (12
dimensional complex vector) on every lattice site. During the step, whichever
variant of the algorithm being used, there is an iteration of the multiplication
of a large ``Wilson spinor'' by the large matrix named ``Dirac operator''
leading to an output ``Wilson spinor''. This part is linear algebra and it is
the most time-consuming part of the algorithm.

The multiplication of the Wilson spinor by the Dirac Operator is mainly
performed in the ETMC code by a routine named ``Hopping\_Matrix'', which
is contributing about 90\% of the total execution time~\cite{Vranas06}.
Wilson spinors as well as the gauge configurations are very large arrays. As we
shall see the major problem to produce efficient computations is to ensure fast
enough data transfer to and from the computing units. It is worth noticing that
the stability of the gauge configuration, during very many iterations of the
multiplication of Wilson spinors by the Dirac operator, {\it allows a
significant reduction of the data to be transferred if one manages to keep the
SU(3) matrices in some kind of fast access memory close to the computing
units}.  This is not easy in general because the gauge matrices constitute
rather large arrays. 

The multiplication of the Wilson spinor by the Dirac
operator is  expressed in formula (\ref{eq:Wilson-Dirac}): the actions of
the Dirac operator involves a sum over quark ``spinors'' ($\psi(i)$) multiplied
by a gluon field ($U_\mu(i)$) through the spin projector.
\begin{equation}\label{eq1}
\chi(i) = \sum_{\mu=x,y,z,t}  \left\{ \kappa_\mu U_\mu(i) \left( I -
\gamma_\mu\right)
\psi(i+\hat \mu) + \kappa_\mu^\ast \, U_\mu^\dagger(i-\hat \mu) \left( I + \gamma_\mu\right)
\psi(i-\hat \mu)\right\}
\label{eq:Wilson-Dirac}
\end{equation}
where $\kappa_x=\kappa_y=\kappa_z=\kappa $ and $\kappa_t=\kappa \exp{i\pi /T}$, 
 $\kappa$ is the "hopping parameter" and the phase $ \exp{i\pi /T}$
expresses the anti-periodic boundary conditions in the time direction.
The gluon field SU(3) matrices are labelled by their starting site and the
space-time direction of the link on which it is defined.

The code we consider contains two variants of the algorithm. The first one named
``full-spinor'' corresponds to the direct application of equation
(\ref{eq:Wilson-Dirac}), while the second named
``half-spinor'' processes via two phases which can be expressed by the
 following set of equations:
First phase (K series):
\begin{equation}
\phi_\mu(i,+) =  \kappa_\mu U_\mu(i) \left( I - \gamma_\mu\right)
\psi(i+\hat \mu)   \qquad
\phi_\mu(i,-)=\left( I + \gamma_\mu\right) \psi(i-\hat \mu)
\end{equation}
 second phase (L series):
 \begin{equation}
\chi(i) = \sum_{\mu=x,y,z,t} \phi_\mu(i,+) + \kappa_\mu^\ast\, U_\mu^\dagger(i-\hat \mu)
 \phi_\mu(i,-).
\label{eq:Wilson-Dirac-half}
\end{equation}
The pros and cons of both variants depend on the architecture and will be
discussed later.

In the next sections we will present studies on numerous architectures
and lattice sizes. Our general goal is the Petaflop as justified in
section 5. Working on one node we have in mind different sublattices
according to different possible decompositions of the full lattice. We
also varied the lattice size in order to highlight the role of the
different architectural components (e.g., cache size, etc).

\section{\label{sec:Singlenode}The Performance of a Single Computing Node}
The performance of lattice QCD on a multiprocessor machine relies heavily on 
the performance of the individual computing nodes. In this section, we will 
start by outlining the architecturally independent attributes of the Hopping\_Matrix 
routine that will interact with the architectures under investigation. We will then 
discuss the performance based on these individual nodes in separate sections.

\begin{figure}
\begin{centering}
\includegraphics[scale=0.44]{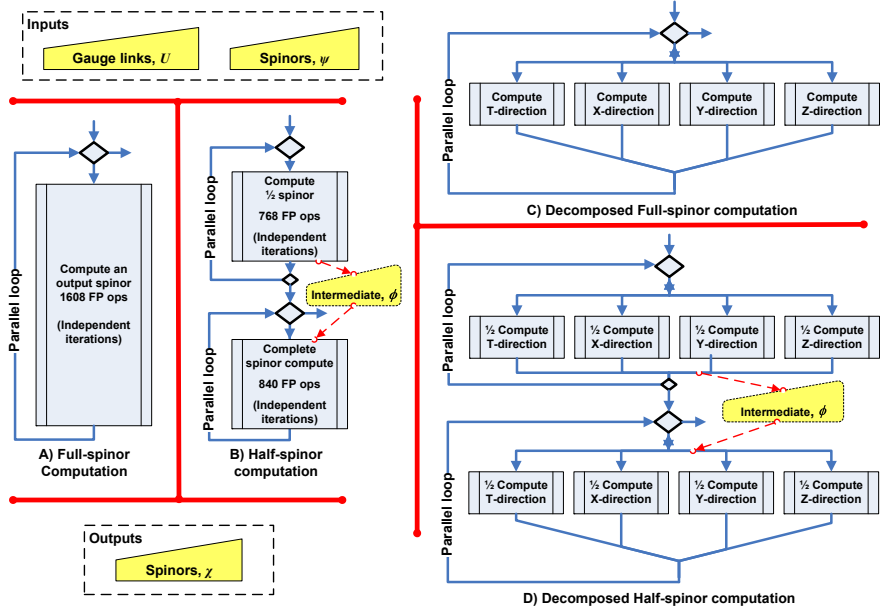} 
\par\end{centering}
\caption{\label{fig:HoppingMatrix}Computation schemes of the Hopping\_Matrix routine based on the full-spinor and the half-spinor versions.}
\end{figure}

\begin{figure}
\begin{centering}
\includegraphics[scale=0.35]{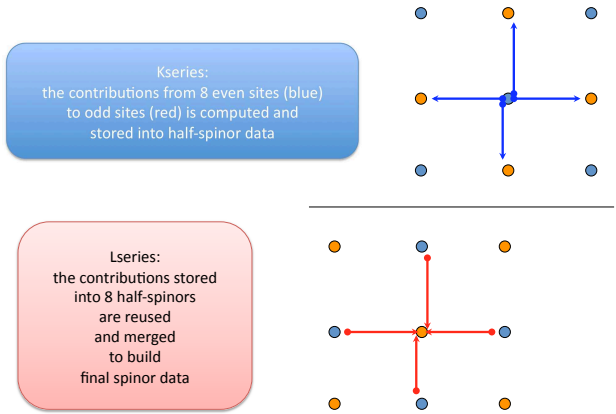}
\par\end{centering}
\caption{\label{fig:Hopping_Matrix-phases}Hopping\_Matrix is splitted into
two phases, Kseries (phase one), then Lseries (phase two) with almost balanced code, data, CPU time.}
\end{figure}

In the Hopping\_Matrix routine, the computation of the spinor involves
1608~\footnote{This number is larger than the commonly quoted  1320 flops per
site, the difference being due to the multiplication by the factor  $\kappa_\mu$ 
in eq. (\ref{eq1}).}
floating-point  operations per lattice site touching 360 floating-point
variables. We focus on the two implementations already mentioned, the
full-spinor and the half-spinor ones. The full-spinor version pulls all the
data needed to compute an output spinor from all surrounding sites. These data
include the gauge field links and the spinors. In each call of the
Hopping\_Matrix routine,  the gauge links show non-redundant regular access,
while reads of the surrounding spinors usually carry redundancy and
irregularity of access because each input spinor appears in the computation of
eight different output spinors. To solve this problem in accessing data, the
half spinor version carries the computation in two phases. In the first phase,
each input spinor is visited once and the computations related to all the
surrounding spinors are pushed to the surrounding spinors in intermediate
half-spinor structures. Writing of the output half-spinors is aligned to
optimize the access pattern in the second phase. In the second phase, the
results of the first phase are used to compute the output spinors. The access of
the half-spinors intermediate structure is more regular. The advantage of the
half-spinor version is that irregular pattern of access is associated with the
writes of the first phase and not with data reads. In most general-purpose
architectures, memory reads are more critical to performance. On the other hand,
the accessed data are increased by about 7\% for the half-spinor version
compared with the full-spinor version.  Figure~\ref{fig:HoppingMatrix} shows
four code variants of the code explored in this study. The computation can be
decomposed into two phases of computation, in the half-spinor version.
Additionally, the computation can be further decomposed based on the number of
space directions. Figure~\ref{fig:Hopping_Matrix-phases} depicts the two main
phases of the  half-spinor computation that allow friendlier cache behavior on
processors  with cache hierarchy.

For processors with normal caches, \emph{e.g.} Itanium and Pentium,
we will focus on the half-spinor version because of its performance
advantage, for computing nodes using accelerators we will explore both
techniques.

\begin{figure}
\begin{centering}
\includegraphics[scale=0.75]{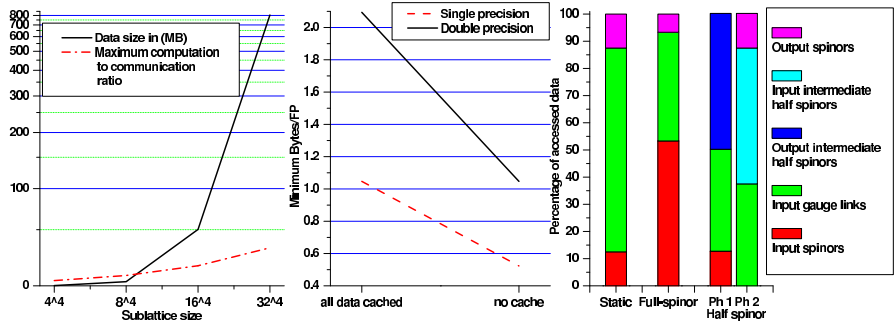}
\par\end{centering}

\caption{\label{fig:Summary-of-Attributes}Summary of the attributes of the
Hopping\_Matrix routine in terms of memory requirements, density of
computation to access and access pattern. }

\end{figure}

The most dominant attribute of Hopping\_Matrix computation that affects
performance is the low computation to memory access ratio, as shown
in Figure~\ref{fig:Summary-of-Attributes}. This analysis assumes
no temporal locality in inter-calls to the Hopping\_Matrix routine.
This assumption is valid taking into account the large footprint associated
with reasonable lattice size, in addition to the alternations in the
computation between multiple input data on consecutive calls to the routine.

This computation to memory access ratio does not exceed 1.05 double precision
floating point operations per byte. This ratio is usually as low as
0.56 FP/byte if the lattice data is not cached. In contrast, reuse of data is related 
to the data size for dense matrix-matrix multiplication, though it is partly 
exploited using blocking due to limited cache sizes. Caching the lattice
data is difficult to achieve because we tend to choose bigger lattice
to mitigate the cost of communication between nodes. The Lattice QCD problem requires
dividing the lattice among many cooperating nodes, which need to communicate
results. To overcome the disparity between the communication
bandwidth and the computational power of the processing node, we tend
to increase the sublattice size per node. The computation grows linearly
with volume while the communication grows linearly with the surface.
Figure~\ref{fig:Summary-of-Attributes} shows that improving (increasing)
the computation to communication ratio linearly requires exponential
growth in the required memory space.

Another worth noting attribute is that the gauge field constitutes
about 75\% of the data accessed (static memory footprint) compared
with 12.5\% for input spinors. At runtime for the full-spinor version, the input spinors
vector represents 55\% of the dynamically accesses data with the least
regular access pattern because of the spin-projection operator. For the half-spinor version, most of the accessed data at runtime belongs to the intermediate data structures carrying the half-spinor data. 

In the following subsections, we present our study first 
on the use of homogenous computing nodes based on Pentium and
Itanium processors, then, on the use of 
heterogeneous nodes where a general purpose processor is assisted by a
special accelerator to speedup floating point computations. We
specifically present our study on the use of Nvidia GPU assisting an
Intel Xeon processor and the use of the IBM Cell BE.

\subsection{Baseline Code - Pentium4}

In order to have a base of comparison, the performance of the HMC/ETMC
code is measured on an Intel Xeon Prescott processor at
3.2GHz with 16KB L1 and 1MB L2 (using one core).


The HMC/ETMC code comes with an optimized version for SSE SIMD
instructions. These are vector instructions able to perform 2 double
precision operations in one single instruction issue. As a matter of
fact the use of SIMD SSE Pentium instructions is explicit in the
code, by the use of dedicated intrinsics. It basically addresses the
computations on the spinor vectors (complex vectors of size 3).

\begin{figure}
\begin{minipage}[b]{0.48\textwidth}
\includegraphics[width=\textwidth]{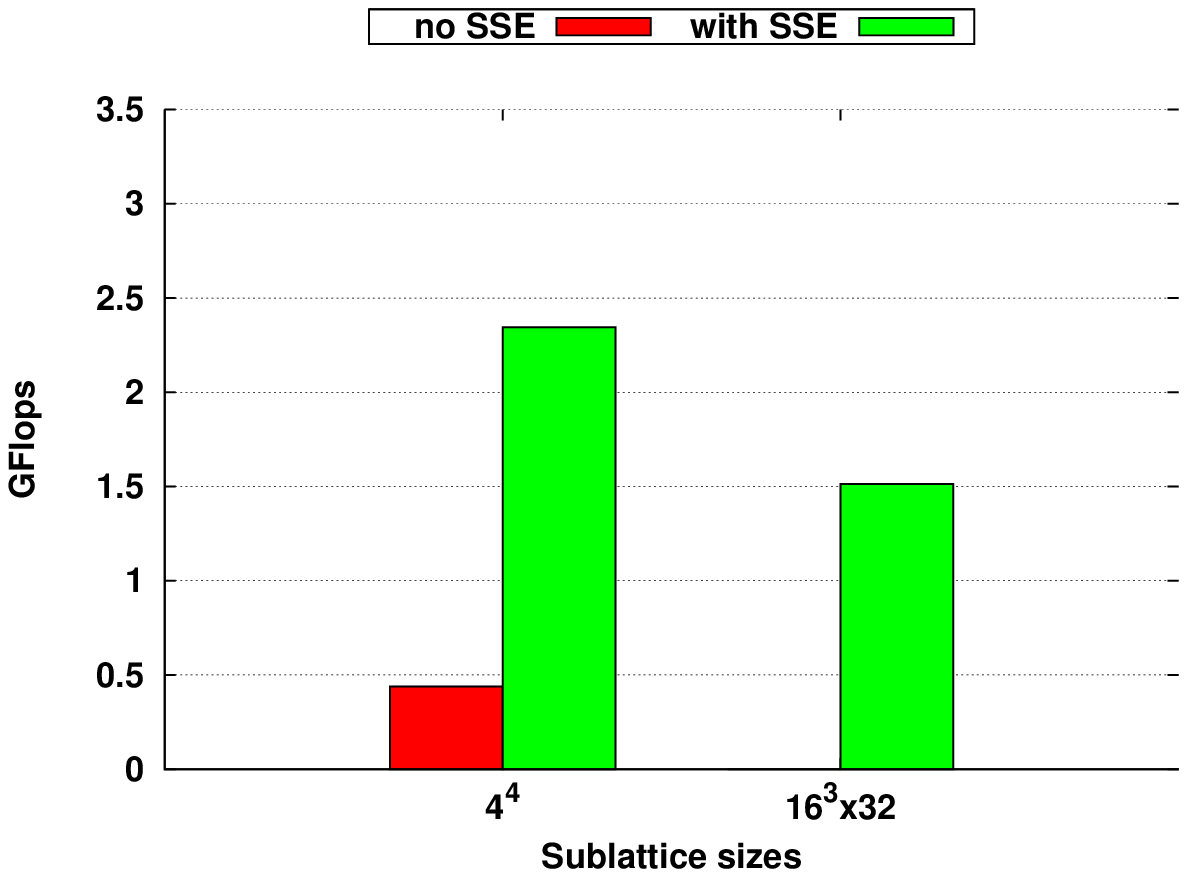} 
\caption{\label{fig:pentium}Performance in GFlops of Hopping\_Matrix on 3.2GHz Pentium4 Prescott, with and without SSE and for different lattice sizes.}
\end{minipage}
~~~~~
\begin{minipage}[b]{0.48\textwidth}
\includegraphics[width=\textwidth]{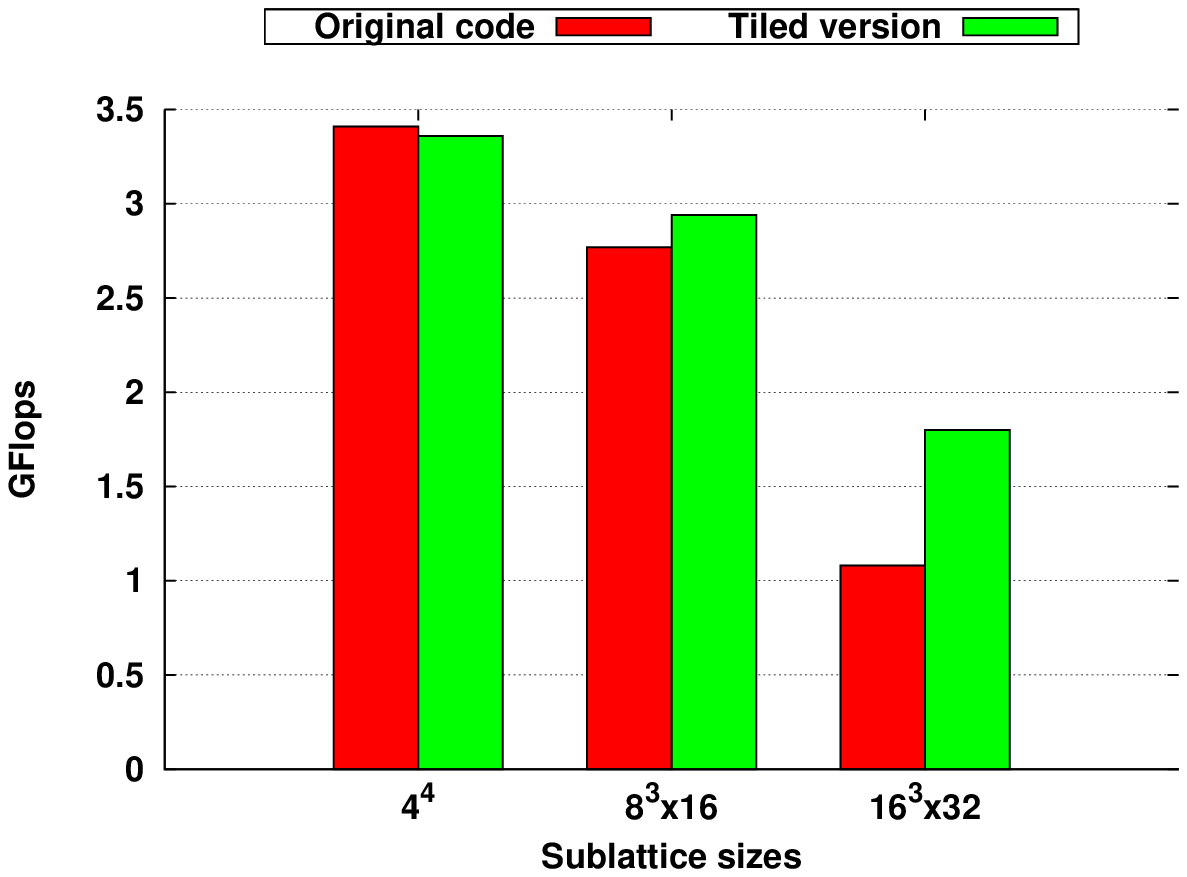}
\caption{\label{fig:perfia64}Performance in GFlops of original Hopping\_Matrix
 and tiled version on 1.6GHz Montvale Itanium2 for different lattice sizes.}
\end{minipage}
\end{figure}

Performance have been measured on two input data sets. The first one
is a lattice of size $4^4$. The second one is a lattice of size $16^3
\times 32$. Performance results are described in Figure
\ref{fig:pentium} for two versions, with and  without SSE.
First, based on the number of floating point operations at each
lattice site (see Section~\ref{sec:Singlenode}) - 1608 operations per
site -, the clock cycle counter and the frequency of the Pentium
(3.2GHz), we found that the speed of the original code is about 2.3
GFlops for the $4^4$ lattice and 1.5 GFlops for the $16^3 \times 32$
lattice when using the SSE instructions. This means that the original
code Pentium version is already highly optimized (peak performance
is 6.4 GFlops in double precision). The better
performance for the smaller lattice is probably due to data reuse that
cannot be exploited so well with the large size.

\def\DO    {{\bf do}}
\def\FOR   {{\bf for}}
\def\WHILE {{\bf while}}
\def\IF    {{\bf if}}
\def\THEN  {{\bf then}}
\def\ELSE  {{\bf else}}
\def\ENDIF {{\bf endif}}
\def\ENDFOR {{\bf endfor}}
\def\DO    {{\bf do}}
\def\ENDDO {{\bf enddo}}
\def\LOOP    {{\bf loop}}
\def\ENDLOOP {{\bf endloop}}
\def\ASM    {{\bf asm}}
\def\ENDASM {{\bf endasm}}
\def\FOR   {{for}}
\def\WHILE {{\bf while}}
\def\ENDWHILE {{\bf endwhile}}
\def\AND   {{.and.}}
\def\PRAGMA{{\bf \#pragma}}


\subsection{\label{sec:itanium}Itanium Architecture}
We evaluate performance of the original HMC code on an Intel Itanium
Montvale processor at 1.67 GHz, with 256KB L2 and 12MB L3 (on a single
core). 


 In the original version, \texttt{Hopping\_Matrix} runs one loop for
each of the two half-spinor computation, one loop for all directions
on the odd sites, then one loop for all directions on the even sites
(Figure
\ref{fig:HoppingMatrix}.B).
The original code suffers from two main deficiencies on Itanium
architecture:
\begin{itemize}
\item Analysis of the compiler generated assembly code~\cite{maqao}
 shows that the compiler has difficulty optimizing the whole basic
 block of the loop. Too many instructions and a too high register
 pressure prevent the compiler for instance to software pipeline the
 loop. This has a high impact on Itanium architecture.  
\item While
 some data (in particular the gauge fields) are reused through the
 computation, the size of the volume prevents data from staying in
 cache between two uses.
\end{itemize}
Note that there is no SIMD code for Itanium, unlike for Pentium, the
code considered is plain C code. This leads to two transformations:
each loop of the two phases is tiled so that data within a tile stays
in cache, and the tiled loop is split for all directions of
computation in order to enhance the quality of compiled code.  Figure
\ref{fig:HoppingMatrix}.D shows the structure of the resulting code:
each of the two parallel loops are tiled and the half-computations
corresponding to each direction within each of these loops are
executed sequentially.

Figure \ref{fig:perfia64}, on the left, shows performance of the two
versions w.r.t. lattice sizes. For a small lattice of size $4^4$, all
the data fit in L2 cache, tiling only introduces overhead and
performance reaches $3.2$ GFlops (peak performance is at 6.4
GFlops in double precision). For a medium size lattice of size $8^3*16$, the data is still
in L3 but no longer in L2. There is a light performance improvement of
the tiled version, that nearly reaches $3$ GFlops. Finally, when the
lattice is too large even for L3 cache, 
the tiled
version outperforms 
the original code by $60\%$. The
effect of tiling reduces the impact of L3 cache misses but as the
reuse factor is low (between 2 and 3) and does not grow with the
lattice volume, memory accesses still drive the performance for large
enough lattices.


The overall performance gain for the whole HMC code reaches $40\%$
speed up for a $16^3*32$ lattice. The best tile size for the
architecture has $128$ iterations and corresponds to a maximum usage
of the cache hierarchy.


\subsection{\label{sec:gpu}Computing Node Based on CPU Assisted by a
  GPU Accelerator}

\begin{figure}
\begin{centering}
\includegraphics[scale=0.8]{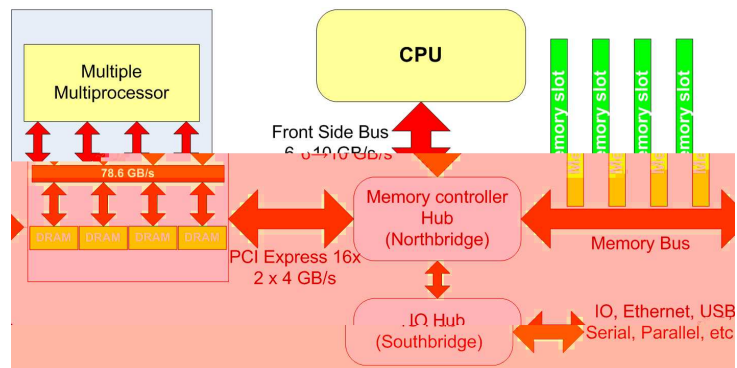} 
\par\end{centering}

\caption{\label{fig:GPU-CPU}Computing node based on a CPU assisted by a GPU
accelerator.}

\end{figure}

The use of GPUs for scientific computing is currently under investigation by 
many research groups and gives very promising results for many scientific 
applications~\cite{egri06}. We investigated the use of NVIDIA G80 GTX graphic 
card GPU board hosted on a server of quad-processor Intel Xeon processor 
at 1.86GHz, 4MB L2 cache.
The NVIDIA G80 GTX GPU is composed of 16 multiprocessors that are
interconnected to banked DRAMs through an impressive bandwidth of
78.6 GB/s. Figure~\ref{fig:GPU-CPU} depicts the layout of connecting
a GPU accelerator to a CPU through a PCI Express bus.

The parallelization process for GPUs is traditionally done based on
vendor compiler. Expressing problem is facilitated with the
advent of general-purpose programming technology, such as CUDA~\cite{NVIDIA_CUDA_10}
by NVIDIA~\cite{NVIDIA_GPU}. Even though parallelization is done
through the compiler, the programmer carries the responsibility of
transforming the code  in a way that enables efficient
parallelism. A must-do transformation is to remove control-flow instructions
whenever possible. For control-flow variables with limited outcomes,
lookup tables can be used or redundant computation in conjunction
with masking to introduce control-flow free code. These techniques
can prove effective in assisting the generation of SIMD operations.

In our implementation, we explored the effect of work granularity on
performance. The granularity one thread impacts on the resources
allocated to this thread, in particular concerning the number of
registers
 (8192 for NVIDIA 8800 GTX). 
Apparently, the Cuda compiler tries not to reduce the amount of
parallelism below 64 threads, assigning at most 128 physical registers
per thread.
Among alternatives explored, we tried the half-spinor version and the
full spinor version. For both versions, either each thread carries the
responsibility of the whole computation of an output spinor
(coarse-grained implementation) or the computation is divided among 16
threads of computation, based on the number of the dimension of the
space.
Each thread iterates through multiple sites of the output spinor array.
The coarser the thread computation the more the stress on the resources 
because more resources are needed to reduce the pressure on memory.
Given that the memory access latency on GPU in the range of 400-600
cycles, it is necessary to reduce the memory access frequency, especially
since the caching within the GPU is severely limited in size.

\begin{figure}
\begin{minipage}[t]{0.48\textwidth}
\includegraphics[width=\textwidth]{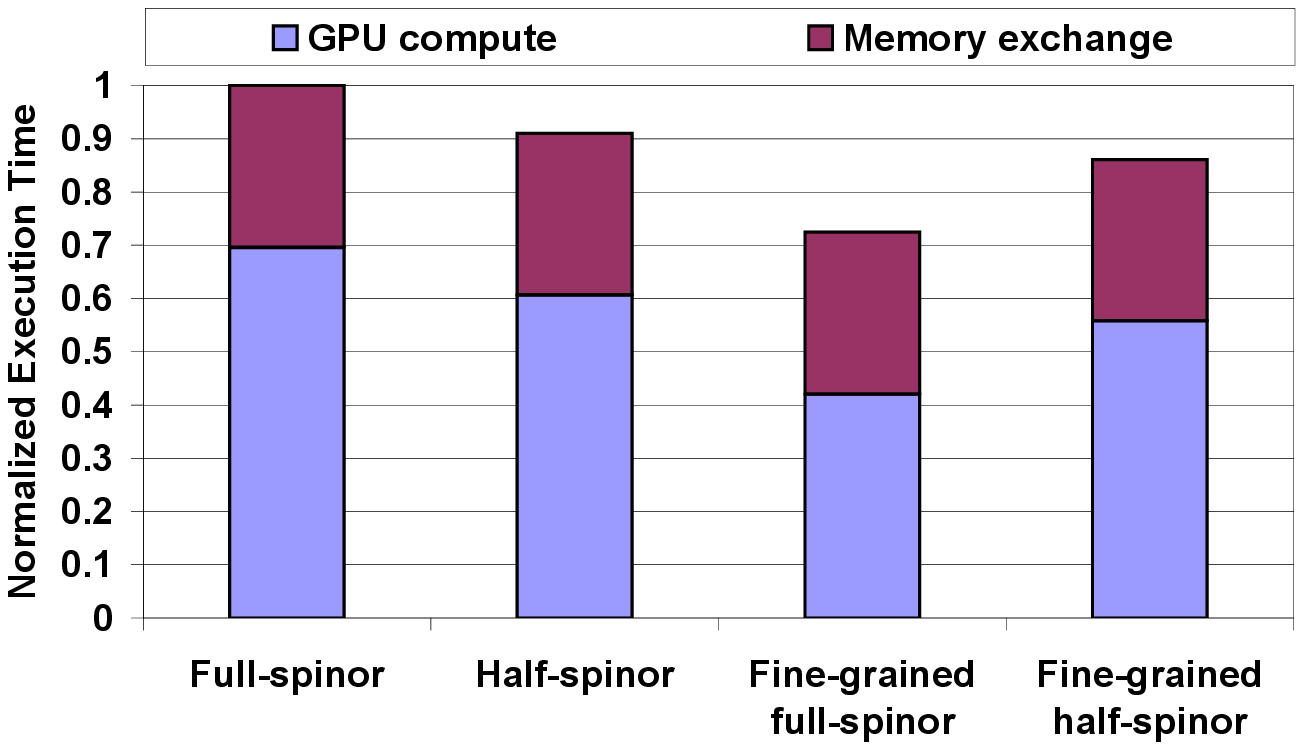} 
\caption{\label{fig:GPU_granularity}The effect of granularity on performance.
Number of threads per multiprocessor is fixed to 64 threads. Fine-grained
versions use 16 threads to carry out the computation of a spinor.}
\end{minipage}
~~~~~
\begin{minipage}[t]{0.48\textwidth}
\includegraphics[width=\textwidth]{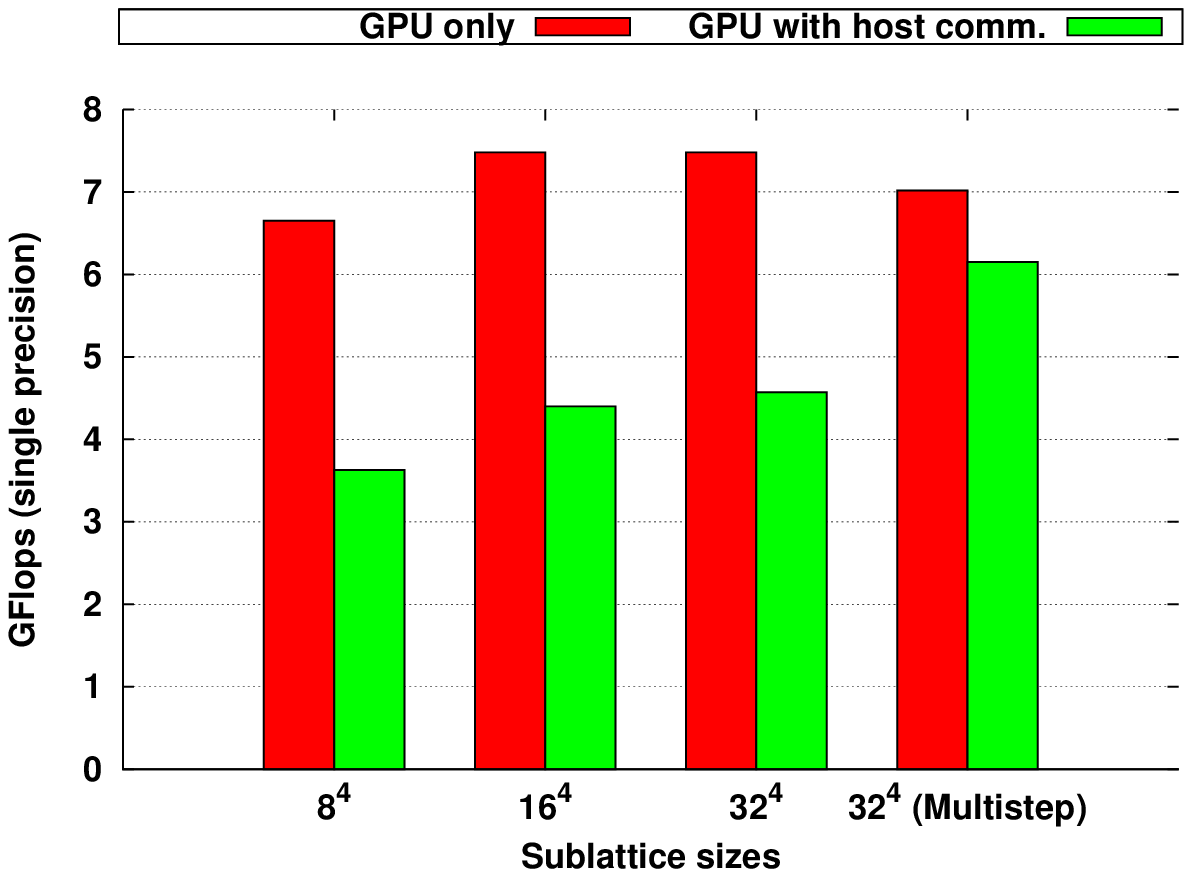} 
\caption{\label{fig:GPU-Performance-Scaling}Performance scaling (in single precision) of the computation for multiple sublattice sizes.}
\end{minipage}
\end{figure}

For Hopping\_Matrix computation, we noticed the less the granularity
of the work assigned to the thread the better the performance
achieved.  Figure~\ref{fig:GPU_granularity} shows the effect of
granularity choice on performance. Number of threads per
multiprocessor is set to 64 (higher number fail to launch for
coarse-grained tasks because of the excessive resources required). We
experimented the two versions of the computation half-spinor and
full-spinor, discussed in Section~\ref{sec:Singlenode}.

For coarse-grained versions, even with the increased memory pressure,
the half-spinor version (two threads per spinor computation) provides
a better performance compared with the full-spinor (one thread per
spinor computation). When the spinor computation are split among 16
threads (of fine-granularity) for both the half-spinor and the full-spinor
techniques, the full-spinor version become better than the half-spinor
version because the former has less frequency of accessing the memory.

The bandwidth of the host CPU memory to the GPU memory has critical
impact on performance. 
Although, the gauge field is
constant across iterations (need not to be exchanged), the input and
the output spinors constitute 25\% of the data accessed in the computation.
Moving these spinors data back and forth between the GPU and the
host processor is of significant cost on performance. The low density
of computation compared with exchanged data causes the communication
overhead between the GPU and CPU to mount to 40\% of the total execution
time even for a large lattice size of $32^{3}\times32$.

To improve the ratio of floating point operations to data exchanged,
we allocated some of the arrays that are used to hold the intermediate
spinor computation to the GPU memory. Using this technique, we reduced
the data exchange frequency to one fourth and the contribution of
data communication to total execution time is lowered to 11\%. This
technique requires communicating the computed spinors less frequently
in a multi-node implementation.

%



Figure~\ref{fig:GPU-Performance-Scaling} shows that increasing the
sublattice size improves the performance up to a certain extent. Using
intermediate spinor arrays to do multiple step of the Hopping\_Matrix
reduces the spinor memory exchange overhead between the CPU memory
and the GPU memory. The performance achieved is about 6.2 Gflops in
single precision.
The efficiency of Lattice QCD computation on GPUs is in the range of
3-4\% of the peak performance because of the low reuse of data and the
complexity of the data access pattern that increases conflicting
accesses of the GPU memory banks.

\subsection{\label{sec:cell}Computing Node Based on the Cell BE}
On the Cell BE, we explored again both the half-spinor and the
full-spinor implementations of the Wilson-Dirac operator. Two data
layouts are considered: small lattices that can fit in the local store
and large lattices that are stored in the main memory and DMAed to the
local store in chunks.

\begin{figure}
\begin{centering}
\includegraphics{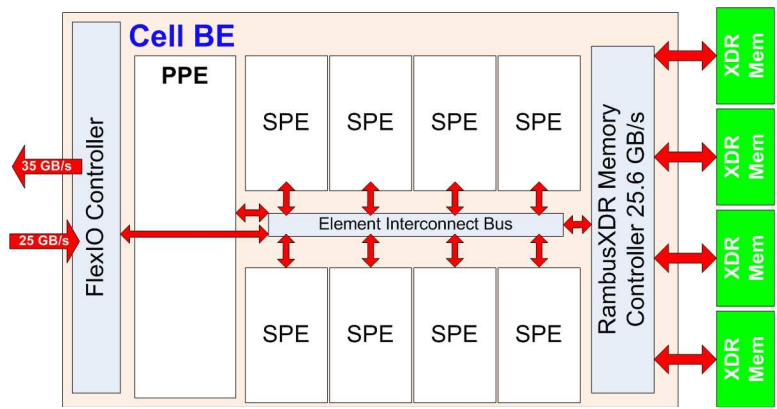} 
\par\end{centering}

\caption{\label{fig:Cell_BE}Computing Node based on accelerator on chip, represented
by IBM Cell BE.}

\end{figure}

Cell broadband engine (BE) is a unique architecture in integrating
specialized accelerator processors, called synergetic processing
element SPE, to the main PowerPC based processor. Each SPE has a
limited memory, called local store, large register file of 128 16-byte
registers and a specialized SIMD processing element. The chip
integrates XDR memory controller in addition to FlexIO
controller. This integration leads to a bandwidth to the memory up to
25.6 GB/s. Figure~\ref{fig:Cell_BE} outlines the main
components of the Cell BE.

The Cell BE is known for being difficult to program partly because
of the detailed control it gives to the programmer over memory management
of the different address spaces of SPEs and the main memory. Special
DMA calls are usually required to control data transfers. Transforming
code to perform efficiently in SIMD mode is an additional traditional
obstacle to exploit SPE processors. Relying on compiler is an option
that is yet to mature for this kind of architecture.
The limited local store size and its separate address space add an
additional dimension to the complexity of accessing the data.  As the
data assigned to a computing node will not fit in the local
store, subset of the data needs to be brought to the local store for
processing.

 Two options exist in bringing data: The first divides the sublattice
assigned to the Cell BE into further smaller sublattices with the
possible data exchanges between SPEs. The second option divides the
computation into frames of data, where SPEs do not need to communicate
data.  The first alternative, which will be presented with the
half-spinor implementation in this work, have the potential of
reducing the pressure on the memory bandwidth. On the other hand, it
requires frequent communication between the SPEs and more
synchronization of points.  The second alternative, which will be
presented with the full-spinor implementation, requires less
synchronization and data exchange between SPEs, but may suffer from
some lost opportunities in reusing data accessed by the neighboring
SPEs.  The little reuse of the data in the Lattice QCD computation
encourages making this trade-off.

\subsubsection{Implementation based on Half-spinor version}

All SU3 objects (vectors, matrices) have been transformed into 4-way complex DP
vectors and 4x4 complex matrices to allow for an easy SIMDization using SPU
intrinsics. This is costly in additional flops (2656 instead of 1608) but
allows, beyond direct measurements, a simple grasp on different scenarios
depending on the size of the local store. It is assumed that 4K sites are
located on each SPU and 128K sites on each CELL. 
  Double buffering is always used across these scenarios. 
They are : \\
S1 : very small current local store size, all of Wilson spinor, gauge matrix 
and half spinors have to be moved in 
or out to/from main memory for each site between both phases ;\\
S2 : half spinor will be kept in  the local store memory (or very close to it) ;\\
S3 : the gauge matrix will be kept into LS or around ;\\
S4 : both the half spinor and gauge matrix  can be allocated into local store
 (the 'Golden Cell').\\
The outcome is that scenario S1 demands a bandwith value well above the
available local store to main memory bandwidth (3.2 GB/sec/spu),
leading to a degraded SPU performance (1.8 GFlop/sec/spu instead of 2.4 for
other scenarios).  Scenario S3 is very interesting, even if it does require an
extra effort about local store size increase : it is worth reminding that the
Gauge Matrices remain constant over many calls of the Hopping\_Matrix routine.

\begin{figure}[ht]
\begin{centering}
\includegraphics[scale=0.4]{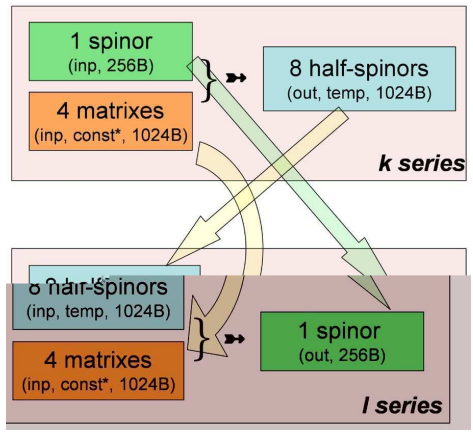}
\par\end{centering}

\caption{\label{fig:data-flow}Site data flows inside of Hopping\_Matrix within
the 2 phases. The input (Kseries) and output (Lseries) spinor indexes
are different (different sites), hence the relevant Gauge Matrices
(site dependent) are also different.}
\end{figure}

\subsubsection{Implementation based on the full-spinor version}

SIMDizing the code requires aligning the data in a way that can be
accessed with the least number of data shuffles. Each spinor is
accessed in eight different contexts (due to the spin projection
operator in Equation~(\ref{eq:Wilson-Dirac}) depending on the space
direction. Each access involves different operations and memory access
pattern for the real and imaginary part of every complex
variable. Unfortunately, SPEs do not support complex arithmetic
instruction set. Dynamic memory accesses of the input spinors
constitute 55\% of the data accessed as shown in
Figure~\ref{fig:Summary-of-Attributes}, while it represents only
12.5\% of the static data accessed. Unfortunately, we cannot fuse
these data statically because the same spinor is accessed in eight
contexts with different surrounding spinors in each case.

\begin{figure}
\begin{minipage}[b]{0.48\textwidth}
\includegraphics[width=\textwidth]{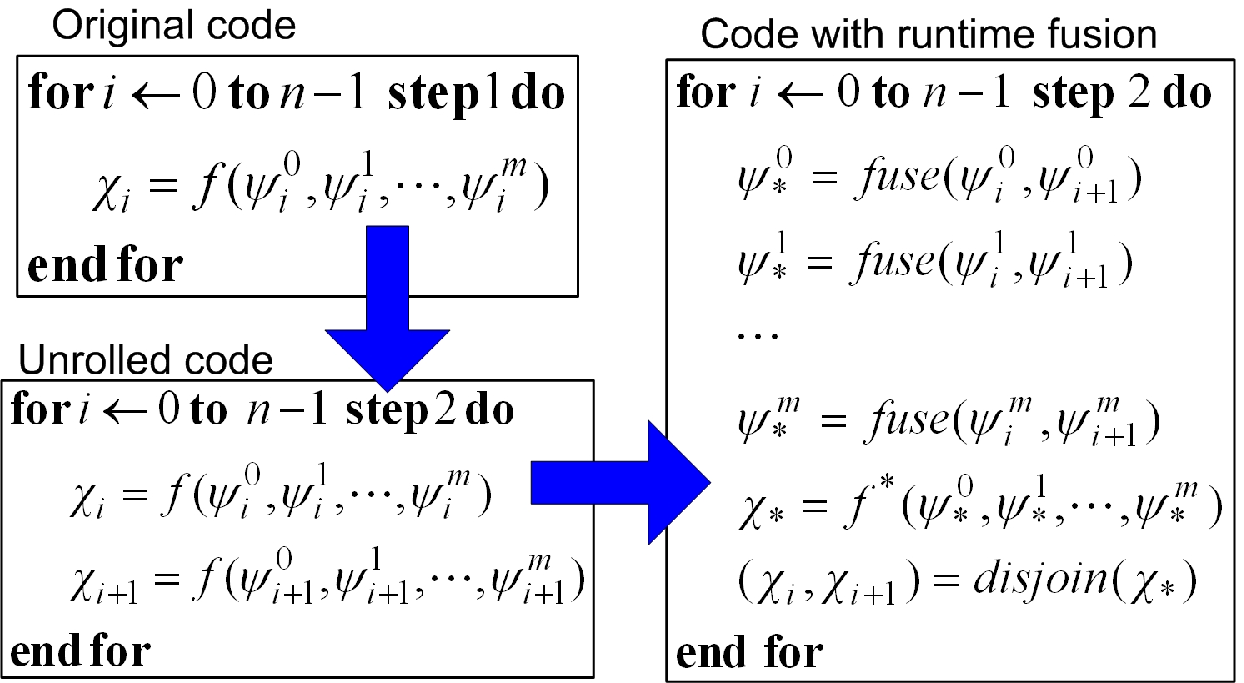} 
\caption{\label{fig:Runtime-fusion}Runtime data fusion technique for the full-spinor version on Cell BE.}
\end{minipage}
~~~~~
\begin{minipage}[b]{0.48\textwidth}
\includegraphics[width=\textwidth]{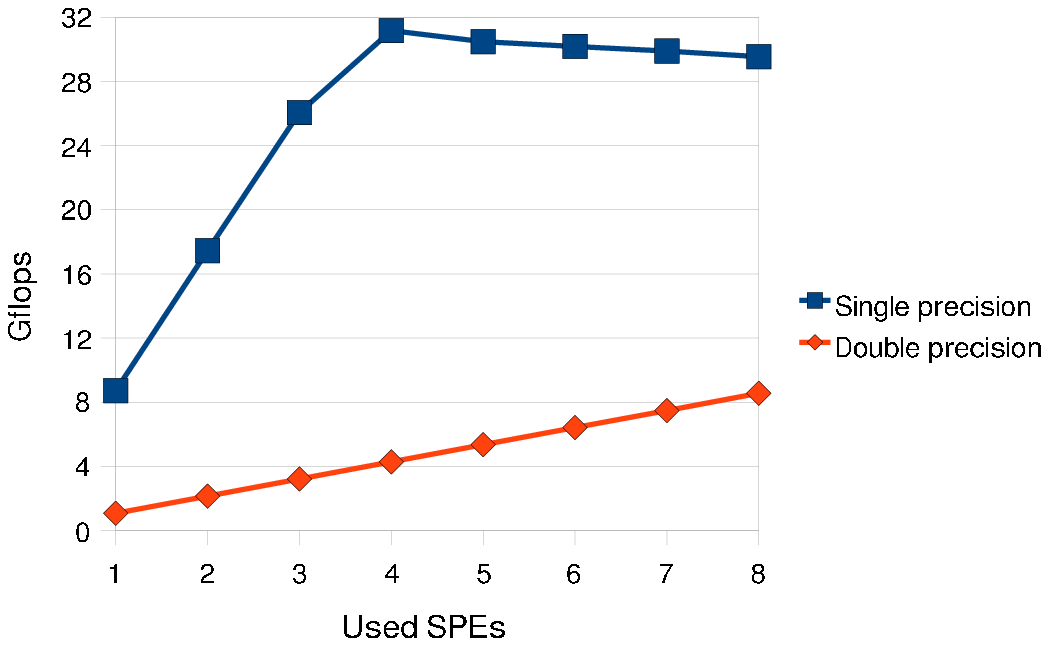} 
\caption{\label{fig:Cell-Perf-Scaling}Performance vs. used SPE for Hopping\_Matrix in single
and double precision.}
\end{minipage}
\end{figure}

To overcome this difficulty, we devise a technique, called runtime
fusion, that fuses the input data used for the computation of multiple
consecutive spinors.  The real parts of these input data are fused
into single register, and similarly for the imaginary part. For
instance a 16 byte register requires fusing the computation of two
output spinors of double precision or four single precision output
spinors. Figure~\ref{fig:Runtime-fusion} shows the layout of the
runtime data fusion technique. Runtime fusion merges the computation
of unrolled loop, thus grouping the data of similar access pattern
into 16 byte words. The result of the computation is then scattered
back into multiple spinors results.  Cell BE allows such technique
because of the large register file. Almost 6~KB of data are touched
during the computation of a group of two output spinors in double
precision.

This technique leads to performing the Hopping\_Matrix routine with 80
Gflops of single precision computation and 8.7 Gflops of double
precision computation. Double precision is not optimized in the
current generation of Cell BE, but PowerXCell 8i with eDP carries an
optimized engine for the double precision that is capable of 50 Gflops
for the Hopping\_Matrix routine.

Realistic lattice size needs to be stored in the main memory and be
retrieved in pieces for processing. The computational power for single
precision Hopping\_Matrix would require 48 GB/s of the memory, far
beyond the 25.6 GB/s bandwidth available.
The input spinor is redundantly accessed 8 times during the computation
of the input spinor array. Bandwidth can be saved, if non-redundant
data are brought to the local store memory from the external
memory, then the redundant part is constructed inside the SPE local
store memory. The saving in bandwidth can be 25\% for a sublattice
size of $16^{3}\times16$.

We exploited above attributes, in the pattern of spinor access, to
achieve computation performance for the Hopping\_Matrix of 31.2 Gflops
for single precision and 8.6 Gflops for double precision.
Figure~\ref{fig:Cell-Perf-Scaling} shows the performance achieved
while changing the number of the SPE used. For single precision computation,
four SPE are able to deliver the maximum the chip can afford. In fact
the performance will slow down by 5\% if all the SPEs are used. The
demand of bandwidth of these 4 SPEs mount to 23.5 GB/s, \emph{i.e.},
almost saturating the bandwidth to the external memory. 

The same behavior is expected for the double precision 
with the new PowerXCell 8i with enhanced double precision.

\section{\label{sec:comparison-and-evolution}Anticipated Future Evolutions and Comparisons}
In this section, we will try to discuss the expected performance
evolution for the studied architectures in the future, for both
homogeneous general-purpose core and based on accelerators. Our study
shows that the use of accelerators can greatly help to boost the
computational performance of the main kernel routine for Lattice
QCD. We will try to discuss the most important criterion that will
influence the choice between the studied accelerator architectures,
for Lattice QCD.

\subsubsection*{Expected advances for Pentium/Itanium}
For the end of the year 2008, the next generation of Itanium
architecture processor, Tukwila, and Xeon processor are expected to
integrate a new memory controller, named Common System Interface. This
controller will offer fast point-to-point processor communication and
will have a peak inter-processor bandwidth of (up to) 96 GB/s and a
peak memory bandwidth of 34 GB/s (first processor are expected to have
only a bandwidth of around 24 GB/s). This would then be comparable to
the current memory bandwidth of Cell BE and would improve performance
for out-of-cache lattices. 

The best performance obtained for \texttt{Hopping\_Matrix} is when all
the lattice fits in cache (L2 and L3). For Monvale processor, this
corresponds to lattices up to the size of $8^3\times16$. Tukwila is planned
to have a 30MB shared L3, for 4 cores. Without any change in the
micro-architecture, a sustained 3 GFlops/core would then be obtained
for lattices of $8^3\times32$. Any increase in the future of cache sizes
would help to maintain a high level of performance.

Experimental results for a whole multi-core processor, taking advantage
of multi-core interactions, are still to be obtained. Efficiency for
Pentium/Itanium code on one core is as high as $50\%$ for smaller
lattices (only considering \texttt{Hopping\_Matrix}) but for the whole
code, it is $18\%$. The efficiency on a multicore node of BlueGene/P is
by comparison of $16\%$, i.e. a sustained Gflops performance $2.2$
Gflops/node (4 core/node).

\subsubsection*{Future prospects of the Cell BE}

The double precision computation is improved on the new generation
Cell EDP engine (PowerXCell 8i). Simulation experiments show that the
Cell EDP is expected to deliver 16 Gflops of double precision
computation. Three to four SPE will also be able to saturate the
bandwidth for double precision because no improvement to the bandwidth
to the memory is introduced.

An increase in the local store size can reduce the pressure on the
bandwidth by improving reuse of the data brought to the SPE. The
unused SPEs can be turned off thus saving power. The performance of
Lattice QCD codes on the Cell BE would improve if the bandwidth to the
memory is improved in the future generations of the Cell.  The kernel
routine implementation can saturate up to double the bandwidth for
single precision computation on current generation Cell BE. For double
precision implementation on Cell EDP, the Hopping\_matrix routine can
saturate more than triple the current memory bandwidth (89 GB/s are
needed to observe 50 Gflops of double precision computation on Cell
EDP). If at one point of time these bandwidths are achieved, then
complex arithmetic instructions would be needed to achieve more
performance.

\subsubsection*{Expected advances on the GPU}
So far, most GPUs lack efficient support for double precision
computation. This is to be rectified in the near
future. Exception handling for floating point is also not supported.

The bandwidth of data exchange between the GPU and the memory is in
the verge of doubling. Because of the dependency of performance on
this scarce bandwidth, we do not expect that having multiple GPU
connected to the same CPU northbridge will be an effective solution.
The efficiency of Lattice QCD computation on GPUs is in the range of
3-4\% of the peak performance because of the low reuse of data and the
complexity of the data access pattern that increases conflicting
accesses of the GPU memory banks.  These issues may require further
investigations for better data alignment.

Reliability of the results obtained by GPUs is a major concern. GPUs
historically served graphic applications that require high performance
but also can tolerate some errors at runtime. Certainly, for
scientific computing this unreliability is difficult to rectify at the
algorithmic level. Software solution to unreliability usually results
in loss of performance.

\subsubsection*{Cell vs. GPU performance comparison}

Among the factors dominating the performance that can be achieved by
any computing node for Lattice QCD are the bandwidths to the memory
system, and the programmability of the computing node. The best
bandwidth observed is currently associated with integrating the memory
controller on the die with the microprocessor. The low computation to
memory access ratio makes the performance heavily reliant on the
memory bandwidth, especially for microprocessor cores with SIMD
instruction set.  The current bandwidth to memory winner is the Cell
BE; that is why it delivers promising performance numbers.

The GPU performance is bounded by the low ratio of computation to data transfer:
a large volume of communicated data has to pass through the bounded bandwidth
between the host memory and the GPU memory. Another challenge is that the
irregular pattern of accessing spinors cannot be handled efficiently when the
job of SIMDization is handed to the compiler.  The compute kernel performance
for Lattice QCD usually relies on hand-coded optimizations to achieve the
most out of the experimented architecture. Expressing the problem in a way that
allows efficient compiler SIMDization requires more study.

The low efficiency of computation on GPUs makes the Watt/Gflops ratio as high as
28. In the PowerXCell 8i, Lattice QCD  requires 3 Watt/Gflops for single
precision and about 6 Watt/Gflops for double precision, assuming none of the SPEs
is turned off.

\subsubsection*{Expected performance evolution and the Lattice QCD problem}

The performance of a single node can increase in the future generation
architecture because of the chance of having higher integration on a
single chip. For instance, the future generation Cell BE is expected
to have more SPEs per Cell chip and more multiprocessor on the GPUs,
and more cores per chip for multi-core systems.

Our study leads us to believe that the efficiency of utilizing the
computational resources on any of these future architectures will
continue to be sub-optimal. The Lattice QCD reuse of data is less than
average applications that most manufacturers balance their design for.

Using/designing a computing facility based on commodity computing
components can be used with Lattice QCD given that enough resource
management is explicitly allowed. Explicit management can allow using
resources based on the balance needed for Lattice QCD, for instance by
switch-off computing resources not used because of the memory
bandwidth bottleneck.

Balancing the resources for a computing node, bandwidth to memory, and
communication between nodes, can be achieved based on resource
management rather than special system designs.

Currently, for Lattice QCD, our study shows that we cannot achieve
less than 3-6 Watt/Gflops, meaning multi mega watts for Petaflops
capable machine. The needed performance for Lattice QCD requires
general technological improvement in performance and power consumption
as well as to facilitate micro-tuning to increase the efficiency of
handling the specifics associated with the Lattice QCD computation.

\section{\label{sec:multinode}Multi-node Systems}
The goal of lattice QCD in the coming years is to compute real QCD i.e.
with light quarks possessing the mass they have in nature. This means
typically a pion twice lighter than usual present computations which
implies a length twice larger in physical units. To increase the accuracy
of the continuum limit and to allow calculations with heavy quarks
a typical reduction of the lattice spacing by a factor 2 will be welcome.
This leads to a multiplication by 4 of the lengths in lattice units, i.e.
a scaling factor of 256. Starting from a lattice of $32^3\times64$ we end up
with a $128^{3}\times256$. This is of course only a rough estimate. We need 
to gain more than two orders of magnitude which amounts indeed to a Petaflops 
sustained performance. The present state of the art, on Bluegene/P, with the
baseline code studied in this paper, reaches about 2.2 Gflops per quadri-core
node, i.e. 22 Teraflops for ten racks (10000 nodes). 

\subsection{The Effect of Communication Architecture on Performance}

Simulating Lattice QCD with physically meaningful size requires
the use of a large number of computing nodes. For instance simulating
a $128^{3}\times256$ lattice requires 8192 nodes each
solving a $16^{3}\times16$ sublattice.

The communication between the 8192 nodes is of critical importance
to the performance, especially when the computing node performance
is improved significantly.
Many machines built for Lattice QCD used 3D torus network for connecting
the  computing nodes~\cite{Boyle01,Belletti06}. A current project for QCD
specialized machine,  QPACE~\cite{qpace}, continues adopting this network
topology with computing nodes based on the PowerXCell 8i.

\begin{figure}
\begin{centering}
\includegraphics[scale=0.85]{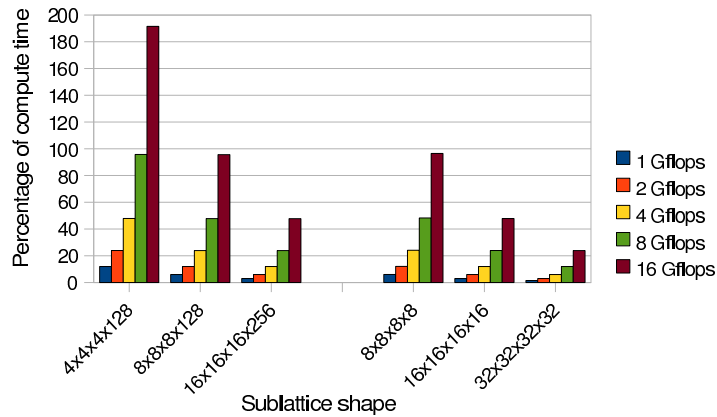}
\par\end{centering}
\vspace{-.5cm}
\caption{\label{fig:Communication-as-percentage}Communication as a percentage
of the compute time for different sublattice shapes and different computation
power of nodes. Sustained link bandwidth is assumed to be 250MB/s per direction. }
\end{figure}

The communication of the Hopping\_Matrix follows the nearest-neighbors
communication pattern. With the large volume of contiguous data communicated,
this communication pattern relies mostly on the link bandwidth to
determine the communication latency.
Assuming a simple model for communication latency given by the equation
$communication\, latency=setup\, time+\nicefrac{data\,
 size}{bandwidth}$.
Then, the communication latency can be computed easily compared with
the computation time. The communication latency depends on the
bandwidth, as a large setup time of 1~$\mu s$ will contribute less than
1\% of the communication latency. In Figure~\ref{fig:Communication-as-percentage},
we present the communication as a percentage of the computation time
for the Hopping\_Matrix routine. We did the computation assuming multiple
performance estimates for the computing nodes range between 1\,Gflops
to 16\,Gflops. For simplicity, we assumed that the computation power
will not vary greatly with the set of sublattice volumes experimented
(carrying 8K to 1M spinors). The sustained link bandwidth is 250MB/s per link,
which is about the expected sustained bandwidth from Blue gene/P interconnection
network (at 55\% of the peak bandwidth).

We have three sublattice volumes each with two structures. For
instance, the sublattice $4^{3}\times128$ is of the same volume as
$8^{3}\times8$, similarly for sublattices $8^{3}\times128$ and
$16^{3}\times16$, and sublattices $16^{3}\times256$ and
$32^{3}\times32$.  The computation to communication ratio is 
proportional to the volume to surface ratio.  The equal edge
sublattices are favored by the bigger computation to communication,
but would require a 4D interconnection network (16
unidirectional links of 250 MB/s sustained). Sublattices with different
link size are what we usually have to embed the four-dimensional
lattice into nodes interconnected with 3D topology.

Assuming 3D interconnection network for a sublattice $4^{3}\times128$,
Figure~\ref{fig:Communication-as-percentage} shows that having a node
of 16 Gflops will lead to a communication that is 1.9 times the
compute time. Increasing the sublattice volume is one solution that
leads to increase the requirement of the memory substantially as shown
earlier in Figure~\ref{fig:Summary-of-Attributes}. For instance to
decrease the communication to 50\% of the compute time, we need to
increase the sublattice to $16^{3}\times256$ (requiring to access to
805 MB in one Hopping\_Matrix call).  We practically try to match the
physical memory to the data accessed in a massively parallel machine
because having virtual memory much larger than the physical memory is
penalized by the expensive IO access, especially for an application
like QCD that streams the data from the memory most of the time, with
little reuse.

Most of the high performance node, like Cell BE and Power6, embed
a memory controlled on the chip and can be connected to a limited
physical memory (usually in the range 0.5 to 2 GBytes).
The communication can be cut to half if we adopt 4-dimensional interconnection
network, assuming preserving the link bandwidth, similar to that of the QCDSP~\cite{Boyle05,qcdsp}, requiring
16 unidirectional links per computing node. 

\section{\label{sec:conclusions}Conclusion}
In this study, we presented the attributes characterizing the main kernel
routine for the Lattice QCD computation. We additionally studied optimizations
and code transformations needed for Lattice QCD on a representative set of
architectures including general-purpose processors, like Itanium,
and the use of commodity floating-point accelerators, such as GPUs and the Cell
BE. 

Most of the optimizations presented in this work target better use of
memory bandwidth, friendlier cache behavior and efficient use of
vector instructions, especially on accelerators. The performance
ranges varied widely, but the use of accelerators provided an appealing
potential especially with the Cell BE. There is also a promising potential 
with GPU accelerators if the above mentioned improvements are introduced.
Neither should one underestimate the potentiality of homogeneous multi-core
architectures with more cores and large caches. The prospects are open and
the foreseeable evolution will be very fast.

The computation to memory access ratio for the Lattice QCD computation is lower
than what is afforded by all the studied architectures and this trend is
expected to continue in the future. Architectures with explicit resource
management can allow more efficient use of the resources.  

We show that the increased performance of computing nodes will
increase the need for having higher performance interconnection
network which was traditionally easily achievable for the Lattice QCD
computation, but will be the critical issue in the future. Algorithmic
improvements such as domain decomposition~\cite{luscher}, which
increase the computation to remote data access ratio, will be welcome.

\section*{Acknowledgements}
We  would like to acknowledge the support of the Agence Nationale de
la Recherche: the project ANR-05-CICG-001 named PARA, and the project 
NT05-3 43577 named QCDNEXT. We also acknowledge the Barcelona Supercomputing 
Center for allowing us access to their Cell clusters.

\end{document}